\documentclass[conference]{IEEEtran}
\IEEEoverridecommandlockouts
\usepackage{cite}
\usepackage{amsmath,amssymb,amsfonts}
\usepackage{algorithmic}
\usepackage{graphicx}
\usepackage{url}
 
\usepackage{textcomp}
\usepackage{multirow}
\usepackage{xcolor}
\usepackage{booktabs}
\def\BibTeX{{\rm B\kern-.05em{\sc i\kern-.025em b}\kern-.08em
    T\kern-.1667em\lower.7ex\hbox{E}\kern-.125emX}}
\begin{document}

\title{BiMoE: Brain-Inspired Experts for EEG-Dominant Affective State Recognition\\
\thanks{This work is supported by the Chongqing Key Project of Technological Innovation and Application (CSTB2023TIAD-STX0015, CSTB2025TIAD-STX0034, CSTB2023TIAD-STX0031, and CSTB2025TIAD-STX0023) and the National Natural Science Foundation of China (Grant No. 62506054).\\
\IEEEauthorrefmark{1} Lin Chen is the corresponding author.}
}

\author{\IEEEauthorblockN{1\textsuperscript{st} Hongyu Zhu, 2\textsuperscript{nd} Lin Chen$^*$, 3\textsuperscript{rd} Mingsheng Shang}
\IEEEauthorblockA{\textit{Chongqing Institute of Green Intelligent Technology, Chinese Academy of Sciences}\\
\textit{Chongqing School, University of Chinese Academy of Sciences}\\
Chongqing, China\\
\{zhuhongyu, chenlin, msshang\}@cigit.ac.cn}
}


\maketitle

\begin{abstract}
Multimodal Sentiment Analysis (MSA) that integrates Electroencephalogram (EEG) with peripheral physiological signals (PPS) is crucial for the development of brain–computer interface (BCI) systems. 
However, existing methods encounter three major challenges: (1) overlooking the region-specific characteristics of affective processing by treating EEG signals as homogeneous; (2) treating EEG as a black-box input, which lacks interpretability into neural representations;(3) ineffective fusion of EEG features with complementary PPS features.
To overcome these issues, we propose BiMoE, a novel brain-inspired mixture of experts framework. BiMoE partitions EEG signals in a brain-topology-aware manner, with each expert utilizing a dual-stream encoder to extract local and global spatiotemporal features. A dedicated expert handles PPS using multi-scale large-kernel convolutions. All experts are dynamically fused through adaptive routing and a joint loss function.
Evaluated under strict subject-independent settings, BiMoE consistently surpasses state-of-the-art baselines across various affective dimensions. On the DEAP and DREAMER datasets, it yields average accuracy improvements of 0.87\% to 5.19\% in multimodal sentiment classification.
The code is available at: \url{https://github.com/HongyuZhu-s/BiMoE}.
\end{abstract}


\section{Introduction}
Emotions play a fundamental role in human cognition, decision-making, and social interaction \cite{2017emotions1,zhu2025ms}. Accurate decoding of affective states from physiological signals is therefore essential for Brain-Computer Interfaces (BCIs), mental health monitoring, and human-centered AI systems \cite{2022tcseption}. 
Electroencephalography (EEG) provides direct, high-temporal-resolution access to neural emotional correlates\cite{2022eeg3}, while Peripheral Physiological Signals (PPS) like Electrocardiograph (ECG), Galvanic Skin Response (GSR), and Respiration (RESP) offer complementary autonomic readings. Capitalizing on this synergy, Multimodal Sentiment Analysis (MSA) based on EEG emerges as a promising paradigm for robust affective state recognition in real-world settings.

\begin{figure}[!ht]
\centerline{\includegraphics[scale=0.62]{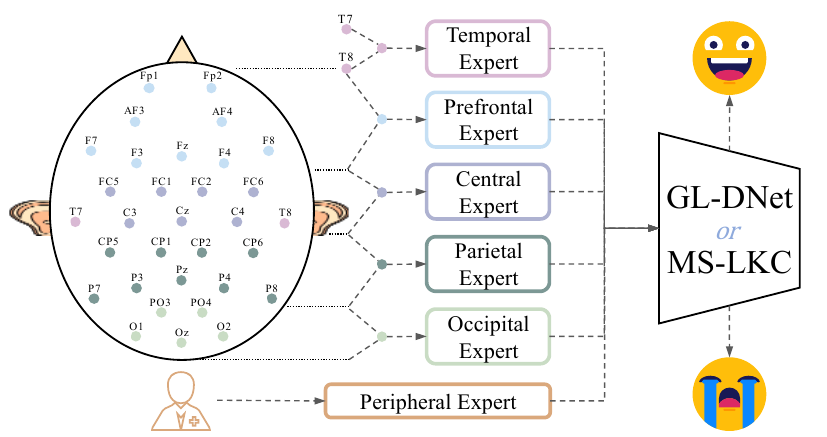}}
    \caption{Expert structure diagram. EEG signals from different regions of the brain and the PPS are assigned to different expert networks for encoding.
    }
    \vspace{-5pt}
    \label{fig1}
\end{figure}

Recent advances in modeling EEG signals have significantly advanced MSA. A prevalent strategy treats multi-channel EEG data as 2D pseudo-images and leverages convolutional neural networks (CNNs) for feature extraction. For example, \cite{li2018hcnns} introduced a hierarchical CNN to capture spatial patterns from EEG-based pseudo-images, while \cite{abidi2024eegcnn2} adopted similar pseudo-image representations for CNN-based encoding. 
Despite their effectiveness, these approaches often neglect the inherent non-Euclidean relationships among EEG channels \cite{hu2025STRFLNet}.
To better model the topological features of EEG, several studies have turned to graph convolutional networks (GCNs). Wang et al.\cite{wang2019PLV} introduced a GCN framework that models phase synchronization between EEG channels using Phase Locking Value (PLV).  Building on this, \cite{hu2025STRFLNet} further incorporated dynamic-static graph topologies characterized by the Phase Lag Index (PLI). In contrast to PLV and PLI, the weighted PLI (wPLI) offers a more robust measure of functional connectivity by mitigating sensitivity to volume conduction and noise \cite{wPLI2019eeg}. However, existing methods often fail to extract and integrate global features with local temporal dynamics efficiently.
Beyond unimodal, recent EEG-dominant multimodal frameworks have explored the integration of PPS to improve performance \cite{2023tacoformer,2024Lanet,2025atpmf,2025MoCERNet,2025cmcler}. Nevertheless, how to effectively fuse EEG with PPS features remains an inadequately addressed problem.

Neuroscientific evidence confirms that emotional processing is mediated by specialized regions of the brain rather than being distributed uniformly throughout \cite{kragel2016decoding}. For instance, the prefrontal cortex is linked to motivational valence and appraisal, while the central cortex contributes to somatosensory integration and arousal\cite{damasio2013nature,2011deap}. Nevertheless, most current modeling of EEG and EEG-dominant MSA methods treat multi-channel EEG as a generic spatiotemporal signal, processing it through black-box networks without incorporating these neuroanatomical priors. 

To address these issues, we proposed BiMoE, a novel mixture of experts (MoE) framework that incorporates neurophysiological knowledge into a neural network (Fig.~\ref{fig1}).
Our approach is structured as follows: First, EEG signals are partitioned according to brain physiological structure and processed by dedicated regional experts. Each expert employs a Global–Local Dual-Stream encoder (GL-DNet), where the multi-layer GCNs with wPLI-based adjacency matrices capture global spatial interactions, and the parallel CNNs extract local temporal features. Simultaneously, PPS are processed by a separate expert utilizing Multi-Scale Large-Kernel CNNs (MS-LKC) to capture their temporal dependencies. Finally, an adaptive router dynamically integrates the outputs of all experts, and the entire model is optimized end-to-end with a joint loss function that balances expert utilization and enhances diversity. To further validate the neuroanatomical alignment, we employ Shapley Additive exPlanations (SHAP) \cite{li2022shap} visualization, which reveals that the model's decision-making process is consistent with established brain region functions, thereby improving the interpretability of BiMoE.

The main contributions are summarized as follows:
\begin{itemize}
  \item  We propose BiMoE, a brain-inspired model whose structure explicitly aligns with the neurophysiology of region-specific affective processing. 
  \item We introduce a MoE framework that employs specialized encoders, including GL-DNet for EEG and MS-LKC for PPS, and integrates a learnable router for dynamic feature fusion, optimized through a joint loss function.
  \item Through extensive subject-independent experiments on the DEAP and DREAMER, BiMoE achieves state-of-the-art performance in both EEG-only and multimodal settings, validating its effectiveness and generalizability.
\end{itemize}

\section{Methodology}
\begin{figure*}[!ht]
\centerline{\includegraphics[scale=0.57]{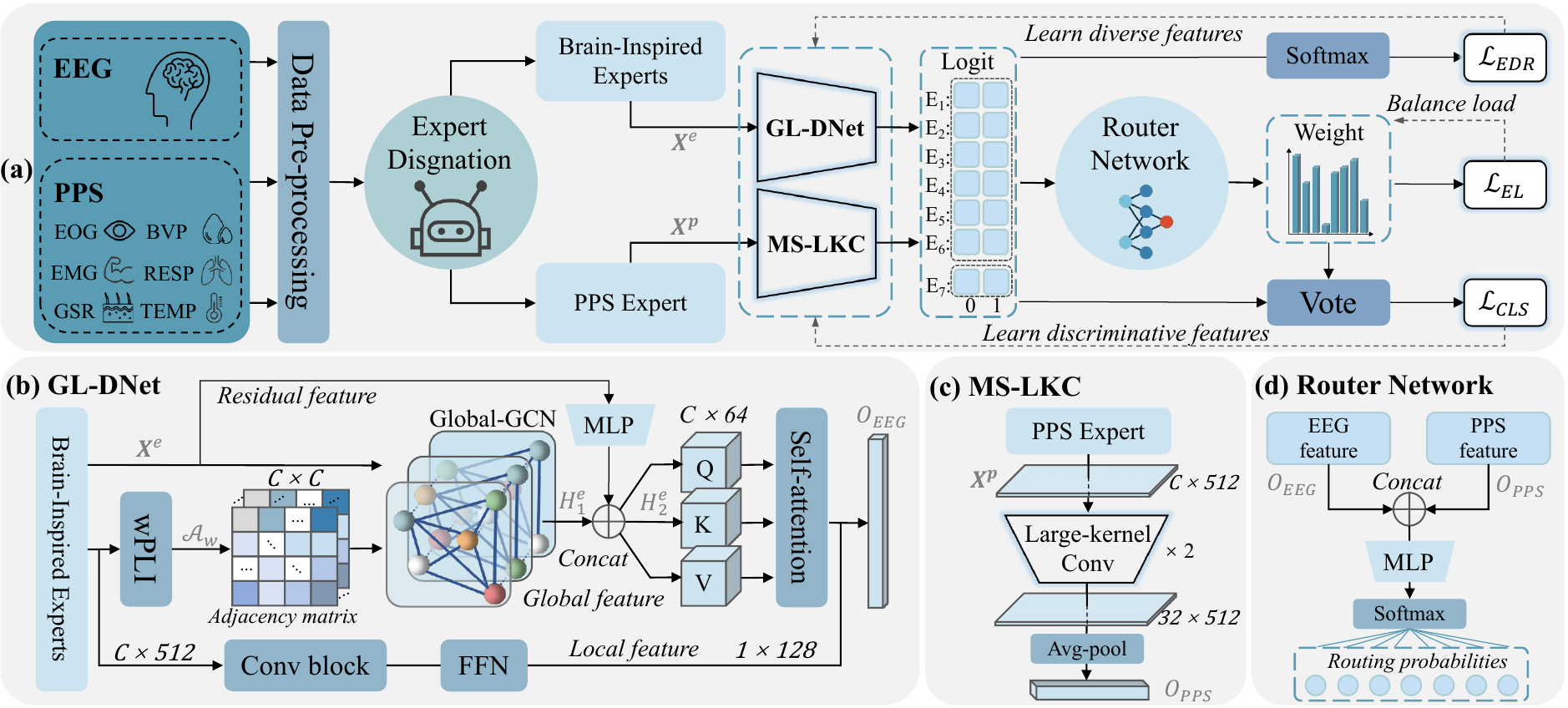}}
    \caption{(a) Overview of the BiMoE framework. BiMoE primarily
comprises: (b) the Global-Local Dual-stream network for EEG processing, (c) the Multi-Scale Large-Kernel Convolutional module for PPS, (d) a Router Network for dynamic feature integration, and a specific joint loss function for optimization.
    }
    \vspace{-5pt}
    \label{fig2}
\end{figure*}

This section details the architecture of our proposed BiMoE framework, as illustrated in Fig.~\ref{fig2}.

\subsection{Task Definition}
Given an MSA dataset comprising $N$ subjects, let $\mathcal{D} = \{(\mathcal{X}, \mathcal{Y} )\}$, where $\mathcal{X}=\{\textbf{X}^{(s)}\}_{s=1}^N$, $\mathcal{Y}=\{\textbf{Y}^{(s)}\}_{s=1}^N$, each pair comprises the multi-channel physiological recordings and the corresponding affective labels for subject $s$. 
For each subject, $ \textbf{X} = (\textbf{X}_1, \textbf{X}_2, \ldots, \textbf{X}_n)$ consists of $n$ samples of physiological signals, while $\textbf{Y} = (\textbf{y}_1, \textbf{y}_2, \ldots, \textbf{y}_n)$ represents the associated ground-truth labels, with each $\textbf{y}_i \in \textit{l}$ corresponding to a multi-dimensional affective rating vector for the $i$-th sample. Where $\textit{l} \subseteq \mathbb{R}^4$ denotes the four affective dimensions: Valence (V), Arousal (A), Dominance (D), and Liking (L).

Under the subject-independent setting for multimodal affective recognition, the objective is to learn a generalizable mapping function $\mathcal{F}: \mathcal{X} \rightarrow \mathcal{Y}$ by training on data from $N-1$ subjects and evaluating on the remaining held-out subject. 

\subsection{Expert Structure}
To align the model architecture with well-established neurophysiological principles, we divided the brain into multiple functional regions, with the EEG signals from each region being processed by its respective dedicated expert.
Formally, let the raw multimodal input of each subject for a trial be $\textbf{X} \in \mathbb{R}^{C \times T}$, where $C$ denotes the total number of physiological channels (EEG and PPS) and $T$ is the sequence length. 

We define a set of brain-region experts, collectively denoted as $\mathcal{E}_{EEG}$=$\{$\textit{prefrontal, central, parietal, occipital, temporal, EEG}$\}$, where the first five correspond to functionally distinct cortical areas and ``\textit{EEG}" processes full-head EEG signals. For each brain-region expert $E_{1-6} \in \mathcal{E}_{\text{EEG}}$, a channel index subset $\mathcal{I}e \subseteq \{1,2,\dots, C\}$ is predefined according to the international 10–20 electrode system \cite{2011deap,2017dreamer}. PPS data $\textbf{X}^p$ are allocated to an additional expert $E_7\in\mathcal{E}_{PPS}=\{peripheral\}$. Thus, the complete set of experts is $\{E_1, E_2, \dots, E_7\}$, where the subscript indicates the expert index.

Fig.~\ref{fig1} illustrates the partition scheme of electrodes employed by the experts. Note that some electrodes (e.g., T7 and T8) are shared across adjacent regions to reflect their dual functional roles. The multi-channel EEG data $\textbf{X}^e$ is then split into region-specific sub-signals:
\begin{equation}
    \begin{aligned}
    \label{eq:1}
    \textbf{X}^{e}=\textbf{X}\left[\mathcal{I}_{e},:\right] \in \mathbb{R}^{C_{e} \times T}, \forall e \in \mathcal{E}_{EEG}.\\
    \end{aligned}
\end{equation}
Where $C_e$ is the number of channels assigned to that EEG expert, thus, the complete data for each subject is represented as $\textbf{X}=\{\textbf{X}^e,\textbf{X}^p\}$. These components are then fed into dedicated expert modules: the proposed GL-DNet for EEG regions and the MS-LKC network for the PPS expert.

\subsection{Global–Local Dual-Stream EEG Feature Extraction}
To capture both the global functional connectivity across brain regions and fine-grained local dynamics within EEG channels, we design the GL-DNet, illustrated in Fig.~\ref{fig2}(b). For each EEG expert $E_{1-6}$, we first compute the \textit{wPLI} between all channel pairs within $\textbf{X}^e$ to quantify region-specific functional connectivity. Formally, for two channels $m$ and $n$ with signals $\textbf{x}^e_m=\textbf{X}^e[m,:]\in \mathbb{R}^T$ and $\textbf{x}^e_n=\textbf{X}^e[n,:]\in \mathbb{R}^T$, we obtain their analytic representations via the Hilbert transform $\mathcal{H}[\cdot]$ \cite{hilbert1970}: 
\begin{equation}
    \begin{aligned}
    \label{eq:2}
    z_m[t]=\textbf{x}^e_m[t]+j \cdot \mathcal{H}[\textbf{x}^e_m[t]]=A^e_m[t]\cdot{e^{j{{\mathrm{\Phi}}_m}[t]}},\\
    z_n[t]=\textbf{x}^e_n[t]+j \cdot \mathcal{H}[\textbf{x}^e_n[t]]=A^e_n[t]\cdot{e^{j{{\mathrm{\Phi}}_n}[t]}},
    \end{aligned}
\end{equation}
where $t=1,2,...,T$ is the time step, $A^e_m[t]$ and $\mathrm{\Phi}_m$ denote the instantaneous amplitude and phase of channel $m$. $j$ represents the imaginary unit. Then, The cross-spectrum $Cs[t]$ is expressed as:
\begin{equation}
    \begin{aligned}
    \label{eq:3}
    Cs[t]=z_m[t]\cdot\mathrm{conj}[z_m[t]],  
    \end{aligned}
\end{equation}
where $\mathrm{conj}(\cdot)$ represents the complex conjugate function.

The \textit{wPLI} between channels $m$ and $n$ is calculated by:
\begin{equation}
    \begin{aligned}
    \label{eq:4}
    wPLI_{mn}=[\frac{1}{T}\sum_{t=1}^{T}{w[t]S[\sin[\Delta\Phi[t]]-S[\cos[\Delta\Phi[t]]]]},
    \end{aligned}
\end{equation}
where the $\Delta\Phi[t]=\mathrm{\Phi}_m[t]-\mathrm{\Phi}_n[t]$ denotes the instantaneous phase difference between $m$ and $n$, obtained via the Hilbert transform. $S[\cdot]$ represents the \textit{Sign} function, and the weighting term $w[t] = Im[Cs[t]]$ corresponds to the imaginary component of the cross-spectrum between the two channels.

After obtaining the \textit{wPLI}-based adjacency matrix $\mathcal{A}_{w}\in [0,1]^{C_e\times C_e}$ for this batch of data, we construct a brain functional connectivity graph and apply the GCN layer to learn topology-aware representations:
\begin{equation}
    \begin{aligned}
    \label{eq:5}
    H^e_1=\mathrm{ReLU}(\mathrm{BN}(\mathrm{GCN}((\textbf{X}^e)^\top,\mathcal{A}_{w}))),
    \end{aligned}
\end{equation}
where \textit{BN} denotes Batch Normalization.

To complement the features with channel-wise contextual refinement, we introduce a residual connection from the $\textbf{X}^e$:
\begin{equation}
    \begin{aligned}
    \label{eq:6}
    H^e_2=\mathrm{Concat}(H^e_1, \mathrm{MLP}(\textbf{X}^e)).
    \end{aligned}
\end{equation}
We then apply a multi-head self-attention mechanism \cite{2017attention} on $H^e_2$ to dynamically reweight channel contributions and obtain the globally attentive features $H^e_g$:
\begin{equation}
    \begin{aligned}
    \label{eq:7}
        head_i=\mathrm{Softmax}\left ( \frac{H^e_2\textbf{W}_i^Q \cdot H^e_2({\textbf{W}_i^K})^\top}{\sqrt{d_{k}}}\right )H^e_2\textbf{W}_i^V, \\
    \end{aligned}
\end{equation}
\begin{equation}
    \begin{aligned}
    \label{eq:8}
        H^e_g=\mathrm{Concat}(head_1, head_2,...,head_h)\textbf{W}^O+H^e_2,
    \end{aligned}
\end{equation}
where $\textbf{W}^{\{Q,K,V,O\}}$ are learnable projection matrices, $d_k$ is the dimension of $K$, the number of heads $h$ is set to 4.

The fine-grained temporal dynamics of the EEG signals are preserved by applying a lightweight convolutional (Conv) block, comprising 1D Conv, \textit{BN} layer, and \textit{ReLU} activation, directly to $\textbf{X}^e$, thus producing the local features. For $E_{1-6}$, the final output of GL-DNet $O_{EEG}$ is obtained by concatenating the global representation and the local features:
\begin{equation}
    \begin{aligned}
    \label{eq:9}
        O_{EEG}=\mathrm{Concat}(H^e_g, \mathrm{FFN}(\mathrm{Conv}(\textbf{X}^e)))\in \mathbb{R}^{{C_e'}\times T},
    \end{aligned}
\end{equation}
where FFN(·) denotes the feed-forward network layer.

\subsection{PPS Representation and Expert Coordination}
For the PPS expert $E_7$, we process its input signal $\textbf{X}^p \in \mathbb{R}^{{C_p}\times T}$ using a multi-scale large-kernel conv block, shown in Fig.\ref{fig2}(c), where $C_p=8$ denotes the number of peripheral channels, e.g., ECG, GSR, RESP. This design captures long-range temporal dependencies in signals without relying on recurrent or attention-based mechanisms.
Specifically, we employ 1D Convs with large kernel sizes (15 and 11) to extract multi-scale temporal patterns, and project them into a unified embedding space using an \textit{MLP} layer:
\begin{equation}
    \begin{aligned}
    \label{eq:10}
    O_{PPS}=\mathrm{MLP}(\mathrm{ReLU}(\mathrm{BN}(\mathrm{Conv}(\textbf{X}^p))) \in \mathbb{R}^{{C_p'}\times T}.
    \end{aligned}
\end{equation}

After obtaining the EEG expert representations $O_{EEG}$ and the peripheral feature $O_{PPS}$, we define the output of each expert as $O^{E_{1-7}}$ and concatenate all expert logits into a unified feature vector $O_f$:
\begin{equation}
    \begin{aligned}
    \label{eq:11}
    O_f=\mathrm{Concat}(O^{E_1},O^{E_2},...,O^{E_7}).
    \end{aligned}
\end{equation}

This fused representation $O_f$ is then fed into a lightweight routing network, as shown in Fig.\ref{fig2}(d), which consists of a two-layer \textit{MLP} followed by a softmax function. This process yields expert-specific weights $W_o\in \mathbb{R}^{7\times 2}$:
\begin{equation}
    \begin{aligned}
    \label{eq:12}
    W_o=\mathrm{Softmax}(W_2\cdot\mathrm{ReLU}(W_1O_f+b_1)+b_2).
    \end{aligned}
\end{equation}
The final prediction probability $y^{pre}$ is computed using Eq.\eqref{eq:13}, which is a weighted sum of all expert logits.
\begin{equation}
    \begin{aligned}
    \label{eq:13}
    y^{pre}=\sum^7_{i=1}W^i_o\cdot(O^{E_{(i)}}) \in \mathbb{R}^{2\times 1}.
    \end{aligned}
\end{equation}

\subsection{Joint Optimization}
The total training objective consists classification loss $\mathcal{L}_{CLS}$, expert load balancing loss $\mathcal{L}_{EL}$ and expert disagreement regularization loss $\mathcal{L}_{EDR}$:
\begin{equation}
    \begin{aligned}
    \label{eq:14}
    \mathcal{L}_{total}=\mathcal{L}_{CLS}+\xi_1\cdot\mathcal{L}_{EL}+\xi_2\cdot\mathcal{L}_{EDR},
    \end{aligned}
\end{equation}
where $\xi_1$ = 100 and $\xi_2$ = 0.05 are hyperparameters used to balance the loss components onto a comparable scale. 

Specifically, to mitigate class imbalance and emphasize hard samples during training, we adopt Focal Loss \cite{2017focal} as the primary classification objective:
\begin{equation}
    \begin{aligned}
    \label{eq:15}
    \mathcal{L}_{CLS}=-\frac{1}{B}\sum^B_{i=1}\alpha_i(1-p^t_i)^\gamma log(p^t_i).
    \end{aligned}
\end{equation}
Where $p^t_i=\mathrm{Softmax}(y^{pre}_i)[y_i]$ is the predicted probability of the true class for sample $i$. We use the simplified configuration with $\gamma=2.0$ and $\alpha=1.0$, allowing the loss to primarily focus on modulating the loss based on sample difficulty. $\alpha$ and $\gamma$ are the class-balancing weight and the focusing parameter.

To prevent certain experts from dominating while others remain underutilized, we introduce expert load loss $\mathcal{L}_{EL}$. First, we encourage all experts to minimize the mean square error $\mathcal{L}_{imp}$ between their utilization rate $U=\frac{1}{B} \sum_{i=1}^{B}W^i_o$ and the uniform distribution.
\begin{equation}
    \begin{aligned}
    \label{eq:16}
    \mathcal{L}_{imp} = \frac{1}{7} \sum_{i=1}^{7} (U_{(i)} - \frac{1}{7})^2.
    \end{aligned}
\end{equation}
Then, we calculate the variance of the utilization rate to prevent any single expert from being overused or ignored. These two parts make up the load loss $\mathcal{L}_{EL}$:
\begin{equation}
    \begin{aligned}
    \label{eq:17}
    \mathcal{L}_{EL} = \frac{1}{7} \sum_{i=1}^{7} (U_{(i)} - \bar{U})^2+\mathcal{L}_{imp}.
    \end{aligned}
\end{equation}

To enhance diversity among experts and discourage redundant learning, we further regularize the model with an expert disagreement loss $\mathcal{L}_{EDR}$. Specifically, we compute the symmetric KL divergence \cite{KLloss} between the softmax-normalized logits $P_i=\mathrm{Softmax}(O^{E_{(i)}})$ of every pair of experts and maximize their pairwise divergence:
\begin{equation}
    \begin{aligned}
    \label{eq:18}
     -\mathcal{L}_{EDR} = -\frac{1}{\binom{7}{2}} \sum_{i=1}^{6} \sum_{j=i+1}^{7} \text{KL}_{{sym}}(P_i \,\|\, P_j),
    \end{aligned}
\end{equation}
where $\text{KL}_{{sym}}(P_i\|P_j)=\frac{1}{2} [\text{KL}(P_i\|P_j)+\text{KL}(P_j\|P_i)]$. By minimizing $-\mathcal{L}_{EDR}$ (i.e., maximizing disagreement), each expert is encouraged to learn different features.
\section{Experiments}

\subsection{Data Preprocessing}
We evaluate our model on two widely used MSA datasets: DEAP~\cite{2011deap} and DREAMER~\cite{2017dreamer}.
Our experiments encompass both EEG-only and multimodal (EEG+PPS) scenarios for comprehensive performance assessment. To ensure fair comparison and reproducibility, we follow the same preprocessing pipeline as established in prior works \cite{2022tcseption,hu2025STRFLNet}.
\paragraph{\textbf{DEAP}} contains 32-channel EEG and 8-channel PPS recorded from 32 participants, with each subject watching 40 one-minute video clips. The PPS modalities include EOG, GSR, EMG, RESP, blood volume pressure (BVP), and temperature (TEMP), all sampled at 512 Hz.
Participants rated affective responses on 9-point scales for Valence, Arousal, Dominance, and Liking. Following \cite{2025MoCERNet}, we binarize ratings at threshold 5, classifying scores $>$5 as ``high" and $\le$5 as ``low".
\paragraph{\textbf{DREAMER}} comprises data from 23 participants viewing 18 videos, featuring 14-channel EEG (128 Hz) and 2-channel ECG (256 Hz) recordings. Participants rated Valence and Arousal on 5-point scales, which we binarize at threshold 3 ($\ge$3 as "high", $<$3 as "low") following \cite{hu2025STRFLNet}.

For both datasets, EEG signals are band-pass filtered (4-45 Hz) using a third-order Butterworth \cite{selesnick2002butterworth} filter to remove artifacts and preserve task-relevant frequencies. All signals are segmented into non-overlapping 1-second windows, followed by channel-wise \textit{Z-score} normalization.

\begin{table}[htbp]
\centering
\caption{ACC of Subject-Independent Experiments On DEAP.} 
\setlength{\tabcolsep}{0.60mm}
\resizebox{1.0\linewidth}{!}{
\begin{tabular}{lccccc}
\toprule
\multirow{2}*{\textbf{Methods}} & \textbf{Published In, } & \multirow{2}*{\textbf{A}} & \multirow{2}*{\textbf{V}} & \multirow{2}*{\textbf{D}} & \multirow{2}*{\textbf{L}} \\
 & \textbf{Year } &   &   &   &  \\
\hline
\multicolumn{6}{c}{EEG only} \\
\hline
CSP\cite{2020CSP} & \textit{ISMC, 2020} & 55.70 & 52.50 & -- & -- \\
FBCSP\cite{2020CSP} & \textit{ISMC, 2020} & 55.31 & 55.23 & -- & -- \\
TSception\cite{2022tcseption} & \textit{TAFFC, 2023} & 63.75 & \underline{62.27} & 64.68* & 65.15* \\
STRFLNet\cite{hu2025STRFLNet}* & \textit{TAFFC, 2025} & \underline{64.18} & 62.12 & \underline{65.08} & \underline{65.83} \\
BiMoE (Ours) & -- & \textbf{65.84} & \textbf{62.96} & \textbf{67.05} & \textbf{67.67} \\
\hline
\multicolumn{6}{c}{EEG + PPS} \\
\hline
TACOFormer\cite{2023tacoformer}$^{\dag}$ & \textit{arXiv, 2023} & 63.07 & 59.12 & 62.74 & 63.62 \\
LANet\cite{2024Lanet}$^{\dag}$ & \textit{INF. FUS., 2024} & 63.22 & 61.94 & 64.85 & 65.72 \\
AT-PMF\cite{2025atpmf}* & \textit{PR, 2025} & \underline{64.87} & 61.54 & 63.98 & 65.40 \\
MoCERNet\cite{2025MoCERNet}$^{\dag}$ & \textit{ACM MM, 2025} & 62.75 & 61.22 & 64.74 & 64.34 \\
CMCLER\cite{2025cmcler}$^{\dag}$ & \textit{MICCAI, 2025} & 64.36 & \underline{62.11} & \underline{66.81} & \underline{67.02}\\
BiMoE (Ours) & -- & \textbf{66.15} & \textbf{63.51} & \textbf{67.68} & \textbf{68.15} \\
\bottomrule
\end{tabular}
}
\vspace{-5pt}
\label{tab:deap}
\end{table}
\begin{table}[htbp]
\centering
\caption{ACC of Subject-Independent Experiments On DREAMER}

\setlength{\tabcolsep}{2.5mm}
\resizebox{1.0\linewidth}{!}{
\begin{tabular}{lccc}
\toprule
\textbf{Methods} & \textbf{Published In, Year} & \textbf{Arousal} & \textbf{Valence} \\
\hline
\multicolumn{4}{c}{EEG only} \\
\hline
BiSMSM\cite{li2022BiSMSM} & \textit{ICANN, 2022} & 63.28 & 60.86 \\
AD-TCNs\cite{2022adtcn}& \textit{CBM, 2022} & 63.69 & 66.56 \\
FCPL\cite{2024fcpl} & \textit{TCDS, 2024} & 65.60 & \underline{67.90} \\
STRFLNet\cite{hu2025STRFLNet}* & \textit{TAFFC, 2025} & \underline{65.76} & 63.25 \\
BiMoE (Ours) & -- &  \textbf{71.55} & \textbf{68.14}  \\
\hline
\multicolumn{4}{c}{EEG + PPS} \\
\hline
CMCLER\cite{2025cmcler}$^\dag$ & \textit{MICCAI, 2025} & \underline{67.70} & \underline{67.06} \\
BiMoE (Ours) & -- &  \textbf{72.89} & \textbf{68.78 } \\
\bottomrule
\end{tabular}
}
\vspace{-5pt}
\label{tab:dreamer}
\end{table}

\begin{figure*}[!ht]
\centerline{\includegraphics[scale=1.1]{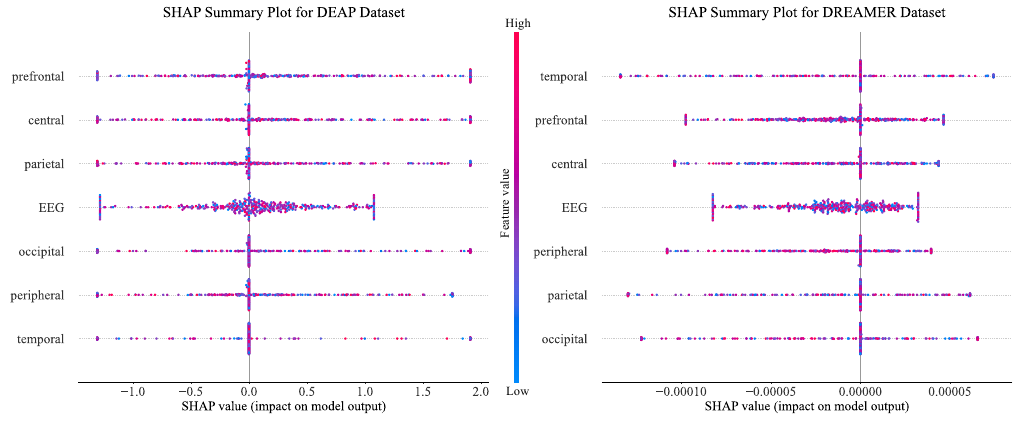}}
    \caption{SHAP summary plots for the DEAP (left) and DREAMER (right) datasets.
    }
    \vspace{-5pt}
    \label{fig3}
\end{figure*}

\subsection{Experiment Settings}
Following a leave-one-subject-out cross-validation setting \cite{hu2025STRFLNet}, we ensure that the training and test sets contain data from different participants.
This process is repeated such that every participant serves as the test subject once, and the final result is reported as the average accuracy across all participants.

Additionally, a sensitivity analysis on the hyperparameters (e.g., $h, \xi_1, \xi_2$) is provided in the Appendix II.A.

All experiments were conducted on a high-performance server equipped with an NVIDIA 3090 Ti GPU, using Python 3.10.18 and PyTorch 2.7.1 with CUDA 12.6 support. To ensure fair comparison, all methods employ identical preprocessing pipelines and data splits.
Ours models are trained using the Adam optimizer with a learning rate of 0.002, weight decay of 5e$‑$4, and a batch size of 128. Training proceeds for a maximum of 100 epochs, with early stopping (patience=25) applied based on validation loss to avoid overfitting. 

\subsection{Verification Performance of BiMoE}
We evaluated BiMoE against comprehensive baselines, including classical\cite{2020CSP,2022tcseption,2023tacoformer,2024Lanet,li2022BiSMSM,2022adtcn,2024fcpl} and state-of-the-art models\cite{hu2025STRFLNet,2025atpmf,2025MoCERNet,2025cmcler}. Our experiments encompass both EEG-only and multimodal (EEG+PPS) scenarios and report average classification accuracy (ACC) for \textit{V}, \textit{A}, \textit{D}, and \textit{L} on DEAP, for \textit{V} and \textit{A} on DREAMER. 

The performance comparisons are summarized in Table \ref{tab:deap} and Table \ref{tab:dreamer}. Results marked with $\dag$ are from our reproduction based on the original descriptions, while $*$ indicates results obtained by running the official code, and others are from the original papers. The \textbf{best} and \underline{second-best} results have been emphasized.
Experimental results show consistent gains by BiMoE. On DEAP (Table \ref{tab:deap}), it improves EEG-only performance by 0.69\% to 1.97\% across \textit{A}, \textit{V}, \textit{D}, and \textit{L} dimensions, and further boosts multimodal performance by 0.87\% to 1.40\%. On DREAMER (Table \ref{tab:dreamer}), notable EEG-only improvements of 5.79\% (\textit{A}) and 0.24\% (\textit{V}) are achieved, which increase to 5.19\% and 1.72\%, respectively, in the multimodal setting.

BiMoE’s brain-inspired architecture uses specialized expert networks corresponding to brain functional regions, providing physiological interpretability. GL-DNet captures global and local EEG features, while MS-LKC extracts multi-scale PPS features with large receptive fields. An automated routing network dynamically allocates expert weights to fuse multimodal representations, optimized via a joint loss function.

\begin{table}[htbp]
\centering
\caption{Ablation experiment on DEAP Dataset}

\setlength{\tabcolsep}{2mm}
\resizebox{1.0\linewidth}{!}{
\begin{tabular}{lcccc}
\toprule
\textbf{Methods} & \textbf{Arousal} & \textbf{Valence} & \textbf{Dominance} & \textbf{Liking} \\
\hline
wPLI $\rightarrow$ PLI & 65.40 & 62.03 & 66.76 & \underline{68.00} \\
w/o expert division & 63.22 & 61.78 & 64.31 & 65.26 \\
w/o self-attention & \underline{65.93} & 61.56 & \underline{67.56} & 67.66 \\
w/o $\mathcal{L}_{EL}$ & 64.42 & 60.85 & 64.28 & 67.61 \\
w/o $\mathcal{L}_{EDR}$ & 65.75 & \underline{62.33} & 67.45 & 67.94 \\
BiMoE (Ours) & \textbf{66.15} & \textbf{63.51} & \textbf{67.68} & \textbf{68.15} \\
\bottomrule
\end{tabular}
}
\vspace{-5pt}
\label{tab:ablation_deap}
\end{table}

To interpret the model decisions, we performed SHAP analysis \cite{li2022shap} on the best subjects for \textit{Arousal} classification (sub 12 from DEAP and sub 18 from DREAMER).
After winsorizing the top / bottom 5\% of expert logits for clarity, summary plots were generated using KernelExplainer with 100 background samples and 400 test samples. 
The results are shown in Fig.~\ref{fig3}, despite variations in feature contributions across subjects and datasets, signals from the \textbf{central} and \textbf{prefrontal} regions consistently dominate, which is fully consistent with the prior knowledge in neurophysiology\cite{damasio2013nature}.

\subsection{Ablation Experiment}
Ablation studies on DEAP validate the contribution of each component in BiMoE (Table \ref{tab:ablation_deap}). Using GL-DNet with raw EEG as a single input instead of brain-region-based expert partitioning reduces performance by up to 3.37\%, highlighting the importance of region-specific modeling. Replacing wPLI with PLI consistently lowers accuracy across all dimensions, highlighting wPLI’s robustness in functional connectivity estimation. Without (w/o) the self-attention in the GL-DNet results in a noticeable drop in \textit{Valence} recognition, underscoring its role in capturing inter-channel dependencies. Disabling the expert load loss $\mathcal{L}_{EL}$ causes a significant performance decline, particularly in \textit{Valence}, highlighting the essential role of balanced expert utilization. Omitting the expert disagreement loss $\mathcal{L}_{EDR}$ also reduces accuracy, verifying that encouraging diversity among experts enhances ensemble robustness. Collectively, these results show that expert structure, wPLI-guided graph learning, self-attention refinement, and multi-objective optimization are essential for effective expert collaboration.

\section{Conclusion}
In this work, we proposed BiMoE, a brain-inspired mixture of experts framework designed to overcome critical challenges in EEG-based multimodal sentiment analysis.
Extensive subject-independent experiments on the DEAP and DREAMER show that BiMoE consistently outperforms state-of-the-art baselines with substantial accuracy improvements across multiple affective dimensions. SHAP analysis further demonstrates that its decisions align with neuroanatomical principles, providing greater interpretability than conventional black-box models.
These advances not only elevate EEG-based MSA but also offer valuable insights for developing interpretable, neurophysiologically grounded architectures in affective computing and broader multimedia applications.

\bibliographystyle{IEEEbib}
\bibliography{references}

\end{document}